%% file: version1.tex
\documentclass[
 reprint, 
 amsmath,
 amssymb,
 aps, 
]{revtex4-1}

\usepackage[utf8]{inputenc}
\usepackage[T1]{fontenc}
\usepackage{times} 
\usepackage{graphicx}
\usepackage{dcolumn}
\usepackage{bm}
\usepackage{bbold}
\usepackage[export]{adjustbox}
\usepackage{xcolor}
\usepackage{caption} 
\usepackage{subcaption} 
\usepackage[colorlinks=true, allcolors=blue]{hyperref} 

\input{my_commands}

\renewcommand{\hide}[1]{}

\DeclareCaptionLabelSeparator{bar}{ | }
\begin{document}

\title{The 1/3 Geometric Constant: Scale Invariance and the Origin of 'Missing Energy' in 3D Quantum Fragmentation
}
\author{Jinzhen Zhu}
\email{zhujinzhenlmu@gmail.com}
\affiliation{%
 Physics Department, Ludwig Maximilians Universit\"at, D-80333 Munich, Germany
}%
\affiliation{%
 Shanghai Artificial Intelligence Laboratory, 129 Longwen Road, Shanghai, China
}

\begin{abstract}
We report the discovery of a universal geometric constraint on the detection of kinetic energy release (KER) in three-dimensional quantum fragmentation. By analyzing the dissociation of localized wavepackets, we demonstrate that the $4\pi r^2$ radial volume element acts as a topological filter that inherently masks a significant portion of a system's energy budget, imposing a fundamental peak-to-mean bound of $R_E < 0.5$. We introduce an invariant scaling law, $\alpha = MQ/\zeta$, and prove that the resulting energy detection ratio is scale-invariant across twelve orders of magnitude, bridging attosecond molecular science and nuclear physics. We identify a universal \textbf{geometric landmark} at $R_E \approx 0.33$, which precisely replicates the 7~eV discrepancy in $H_2^+$ fragmentation. Furthermore, we show that the population of excited-state manifolds and the increase in nuclear localization ($\zeta$) provide a definitive geometric mechanism for the \textbf{spectral broadening} observed across atomic and subatomic scales. Remarkably, the spectral morphology derived from our scaling law aligns with the universal 1/3 energy landmark of historical beta decay, while the high-mass limit naturally accounts for the sharpening of alpha spectra. Our results suggest that ``missing energy'' is often a topological artifact of 3D geometry rather than an exclusive signature of undetected particles. This work establishes a universal master curve for energy reconstruction and identifies a \textbf{``detection crisis''} in highly localized systems, where the true interaction energy becomes effectively invisible to peak-centric calorimetry.
\end{abstract}

\maketitle

\section{Introduction}

A central challenge in experimental physics is the reconstruction of a system's total energy budget from detected spectral distributions. Across twelve orders of magnitude—from the electron-volts of molecular dissociation to the mega-electron-volts of nuclear transitions—a persistent anomaly remains: the primary spectral peak often captures only a fraction of the available interaction energy. Historically, such ``missing energy'' has served as a catalyst for fundamental discoveries, most notably the neutrino hypothesis to resolve the energy deficit in $\beta$ decay~\cite{Ellis1927}.
\begin{figure}
\centering
\includegraphics[width=0.50\textwidth,clip]{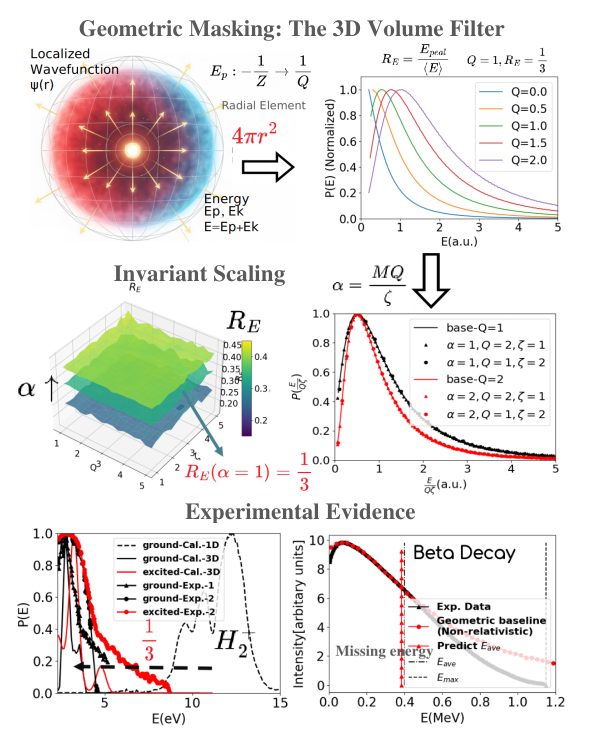}
\caption{
Universal Invariant Scaling and the Geometric Masking of Kinetic Energy. (Row 1) Physical mechanism: The transition from a localized bound state ($-Z$) to a repulsive continuum ($+Q$) is filtered by the $4\pi r^2$ radial volume element, shifting the detected energy peak ($R_E$) significantly below the true system budget. (Row 2) Mathematical framework: The invariant scaling law defines a universal manifold where energy detection ratios depend solely on the coupling of localization ($\zeta$) and repulsion ($Q$). (Row 3) Empirical validation: The 1/3 geometric landmark ($R_E=0.33$) provides a unified explanation for the 7~eV "missing energy" in attosecond $H_2^+$ fragmentation and the historical average-to-endpoint discrepancy in $^{210}$Bi beta decay.
}
\label{fig:summary}
\end{figure}

In this Article, we propose that these disparate anomalies share a common topological origin. As illustrated in our conceptual framework (Fig.~\ref{fig:summary}), we identify a \textbf{Geometric Masking} effect arising from the inherent $4\pi r^2$ volume weighting of three-dimensional space. This weighting suppresses detection probability near the coordinate origin—where potential energy is maximal—systematically shifting the detected kinetic energy release (KER) peak away from the statistical mean (Fig.~\ref{fig:summary}, Row 1).

We demonstrate that this masking is governed by an \textbf{invariant scaling law}
\begin{equation}
\alpha = \frac{MQ}{\zeta},
\end{equation}
which couples the fragment mass ($M$), orbital localization ($\zeta$), and repulsive charge ($Q$). We prove that for any localized 3D wavefunction, this coupling constrains the energy detection ratio $R_E = E_{\text{peak}}/\langle E \rangle$. Specifically, we identify a \textbf{0.33 geometric landmark} (Fig.~\ref{fig:summary}, Row 2)—a universal attractor where the detected peak represents approximately one-third of the total energy budget. Our framework predicts a ``detection crisis'' in highly localized systems: as $\zeta$ increases, $R_E$ decays, rendering the true interaction energy effectively invisible to peak-centric calorimetry.

The predictive utility of this scaling law is evidenced by its alignment with diverse experimental benchmarks (Fig.~\ref{fig:summary}, Row 3). In attosecond science, our model resolves the long-standing 7~eV discrepancy in $H_2^+$ fragmentation, yielding a detection ratio of $R_E \approx 0.32$ that matches modern experimental data~\cite{Zhu2020b, Zhu2021} while accounting for nodal spectral broadening. Remarkably, this molecular landmark is mirrored in the subatomic regime; our distribution provides a robust fit for the historical $^{210}$Bi $\beta$-decay spectrum, where the average electron energy remains tethered to approximately one-third of the endpoint energy~\cite{Ellis1927}. This suggests that the 0.33 ratio functions as a universal \textbf{geometric resonance gate}, where energy transfer is dominated by the maximum spatial overlap of 3D wavefunctions. The persistence of this ratio across twelve orders of magnitude identifies the 3D volume element as a universal arbiter of energy detection. Ultimately, this framework establishes a unified baseline for energy reconstruction, providing a rigorous geometric filter to distinguish intrinsic topological masking from the signatures of new physics.

\section{Theoretical Framework and Numerical Validation}

\subsection{Sudden Approximation and Bohmian Local Energy}
The core of our physical model rests on the sudden approximation, which is valid when the system's Hamiltonian evolves on a timescale $\tau_{\text{pert}} \ll \hbar/\Delta E$. Under this condition, the initial wavefunction $\psi_0(\mathbf{r})$ remains ``frozen'' at $t=0$, as the probability density has insufficient time to reorganize~\cite{Makarov2016Sudden}. We model the fragmentation event as an instantaneous transition from an attractive bound potential to a repulsive Coulomb configuration:
\begin{equation}
    V_{\text{initial}}(r) = -\frac{Z_{\text{att}}}{r} \longrightarrow V_{\text{final}}(r) = +\frac{Q}{r}.
\end{equation}
This framework captures the essential non-adiabatic dynamics of ultrafast processes across scales, from attosecond molecular dissociation to the rapid loss of confinement in nuclear physics (see Sec.~\ref{sec:sudden} for a detailed comment).

To map the initial-state geometry to final energy observables, we utilize the de Broglie-Bohm formulation of quantum mechanics. In this representation, each point in the initial density corresponds to a trajectory with a conserved total energy $E(\mathbf{r})$, defined as the sum of the external repulsive potential and the local (quantum) kinetic energy density $T_{\text{loc}}(\mathbf{r}) = -\frac{1}{2M} \frac{\nabla^2 \psi_0(\mathbf{r})}{\psi_0(\mathbf{r})}$. For a generalized Slater-type orbital $R(r) = N r^{n-1} e^{-\zeta r}$, the analytical expression for the conserved total energy is:
\begin{equation}
    E(r) = -\frac{1}{2M} \left( \zeta^2 - \frac{2n\zeta}{r} + \frac{n(n-1)}{r^2} \right) + \frac{Q}{r}.
\end{equation}

Due to energy conservation, a particle released from position $r$ retains the energy of its initial state. Consequently, the observed fragment energy distribution $P(E)$ is equivalent to that of the initial state, constructed by sampling these local energies according to the 3D radial probability density $\rho(r) = 4\pi r^2 |\psi_0(r)|^2$. Crucially, the $r^2$ volume weighting acts as a \textbf{topological filter}. In highly localized systems (high $\zeta$), the probability density is concentrated at small $r$, where the local kinetic energy is significantly negative due to high wave-function curvature. This geometric weighting ensures that the mode of the distribution ($E_{\text{peak}}$) is systematically shifted to energies lower than the integrated mean $\langle E \rangle$.

\subsection{Comparison with TDSE Benchmarks}
To evaluate the predictive power of this 3D analytical model, we benchmarked the resulting $P(E)$ distributions against full-dimensional quantum simulations using the \texttt{tRecX} Time-Dependent Schr\"{o}dinger Equation (TDSE) solver~\cite{SCRINZI2022108146}. While comprehensive comparisons across different grid parameters are provided in Sec.~\ref{sec:fitting}, we emphasize that our analytical local-energy approach captures the primary experimental signatures---specifically the peak energy $E_{\text{peak}}$---with high precision across both the perturbative and saturation regimes. The consistency between the Bohmian peak predictions and the full TDSE results (see Fig.~\ref{fig:trecx_compare}) confirms that the $4\pi r^2$ volume element, rather than complex dynamical interference, is the dominant driver of the observed energy morphology in 3D Coulomb explosions.
\subsection{Parametric Sensitivity and Spectral Morphology}
\begin{figure*}[htbp]
\centering
\centering
    \begin{subfigure}[b]{0.48\textwidth}
        \centering
        \includegraphics[width=\textwidth, clip]{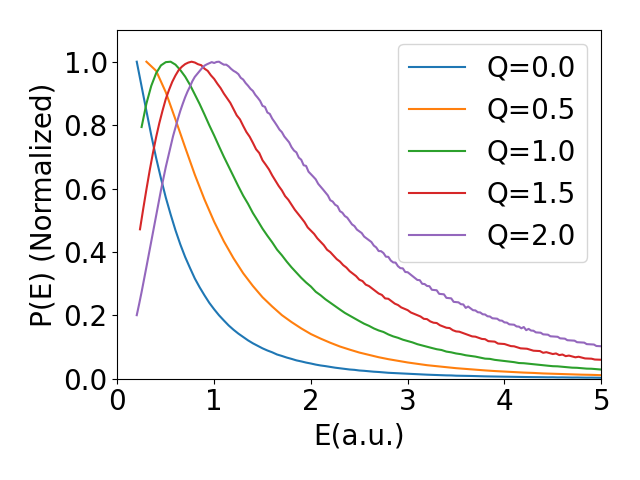}
        \caption{}
        \label{fig:coulomb_test}
    \end{subfigure}
    \hfill
    \begin{subfigure}[b]{0.48\textwidth}
        \centering
        \includegraphics[width=\textwidth, clip]{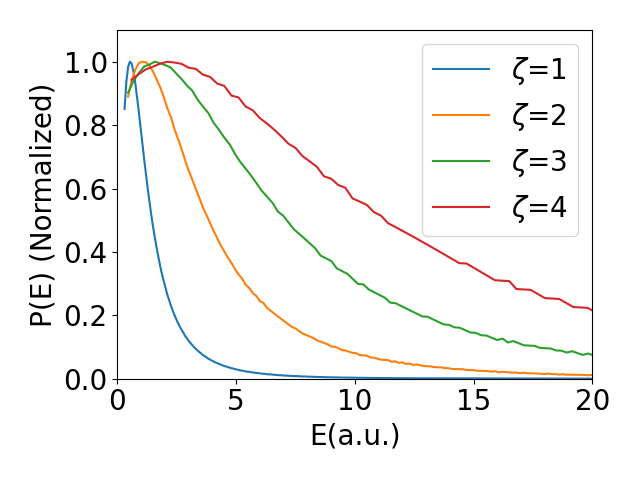}
        \caption{}
        \label{fig:zeta_test}
    \end{subfigure}

    \vspace{12pt} 

    \begin{subfigure}[b]{0.48\textwidth}
        \centering
        \includegraphics[width=\textwidth, clip]{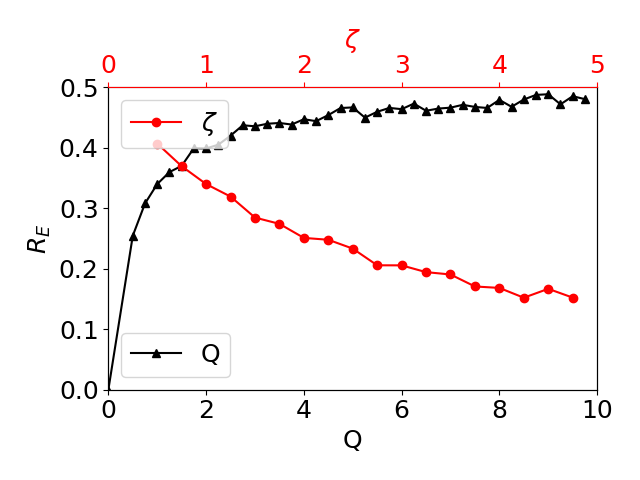}
        \caption{}
        \label{fig:zeta_q_ratio}
    \end{subfigure}
    \hfill
    \begin{subfigure}[b]{0.48\textwidth}
        \centering
        \includegraphics[width=\textwidth, clip]{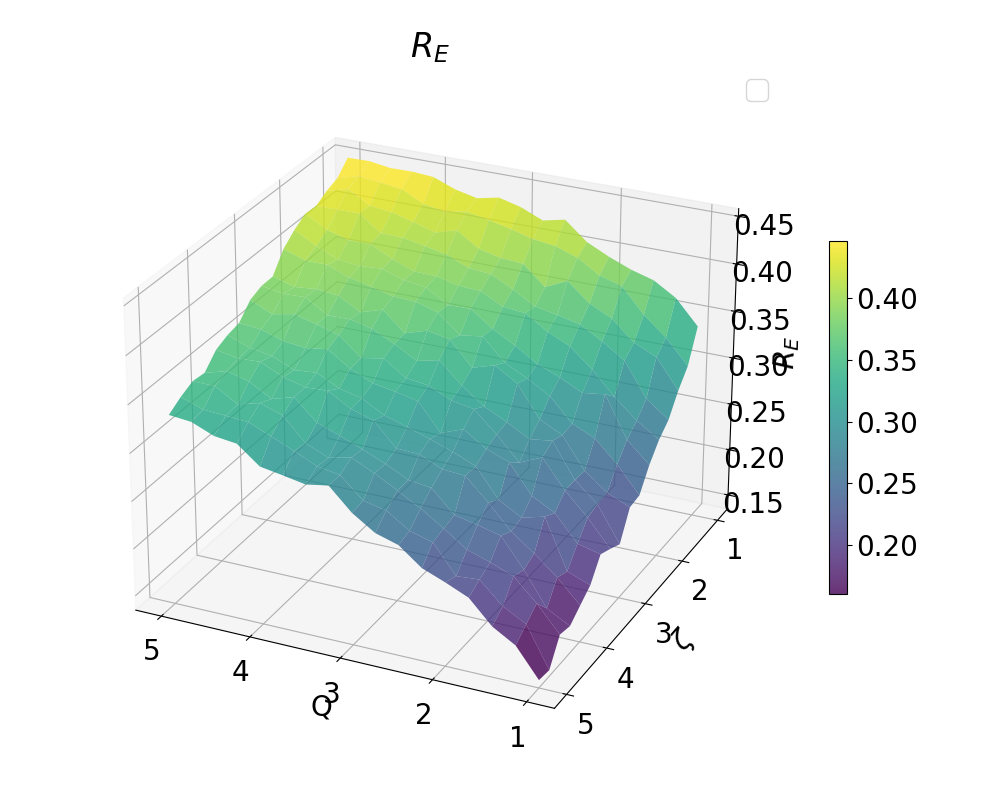}
        \caption{}
        \label{fig:coulomb_zeta_ratio}
    \end{subfigure}
\caption{
\textbf{Parametric evolution and universal scaling of the peak-to-mean energy ratio.} 
\textbf{(a)} Sensitivity of the energy distribution $P(E)$ to repulsive Coulomb charge $Q \in [0, 2]$ for a fixed $1s$ orbital ($\zeta=1, l=0$). 
\textbf{(b)} Evolution of $P(E)$ with orbital exponent $\zeta \in [1, 4]$ at fixed $Q=1$. 
\textbf{(c)} Evolution of the energy ratio $R_E = E_{\text{peak}}/\langle E \rangle$ as a function of $Q$ (at $\zeta=1$, black curve) and $\zeta$ (at $Q=1$, red curve), demonstrating convergence toward $0.5$ in the repulsion-dominated regime and an efficiency drop in the localization-dominated regime. 
\textbf{(d)} Two-dimensional contour map of $R_E$ across the $(\zeta, Q)$ parameter space. The color scale indicates the magnitude of the detection discrepancy, strictly bounded by the 3D geometric masking limit.
}
\label{fig:Coulomb-Zeta-test}
\end{figure*}
To characterize the influence of physical parameters on the detected energy landscape, we systematically varied the repulsive Coulomb charge $Q$ and the orbital exponent $\zeta$ (Fig.~\ref{fig:Coulomb-Zeta-test}). As illustrated in Fig.~\ref{fig:coulomb_test}, increasing the repulsive charge $Q$ induces a linear translation of the distribution $P(E)$ toward higher energies. This shift is accompanied by a moderate spectral broadening, as a larger $Q$ amplifies the potential gradient across the spatial support of the initial wavefunction.

In contrast, the evolution of $P(E)$ with respect to the orbital exponent $\zeta$ reveals a non-linear sensitivity (Fig.~\ref{fig:zeta_test}). As $\zeta$ increases, the wavefunction contracts, simultaneously elevating the local kinetic energy density—governed by the quadratic scaling $T_{\text{local}} \propto \zeta^2$—and shifting the density into regions of steeper potential energy. Notably, the distribution broadens significantly more rapidly with $\zeta$ than with $Q$. This suggests that the internal ``quantum pressure'' of the initial state is the primary driver of spectral dispersion, outweighing the influence of the external repulsive field. These morphological trends confirm that fragmentation spectra fundamentally function as a probe of the initial-state radial geometry rather than a direct measurement of the integrated energy mean.

\subsection{The Energy Detection Ratio $R_E$ and the Detection Crisis}

To quantify the discrepancy between experimentally observable fragment signals and the total energy budget, we define the dimensionless detection ratio:
\begin{equation}
R_E = \frac{E_{\text{peak}}}{\langle E \rangle}
\end{equation}
where $E_{\text{peak}}$ corresponds to the mode of $P(E)$ and $\langle E \rangle$ represents the integrated expectation value of the total energy. This ratio serves as a diagnostic for the efficiency of peak-centric calorimetry in capturing the true system energy. Figures~\ref{fig:zeta_q_ratio} and~\ref{fig:coulomb_zeta_ratio} illustrate the systematic evolution of $R_E$ across the parametric landscape.

\subsubsection{Asymptotic Convergence and Classicalization}
As shown by the black curve in Fig.~\ref{fig:zeta_q_ratio}, $R_E$ increases monotonically with the repulsive charge $Q$, asymptotically approaching a universal limit of $0.5$. In this high-$Q$ regime, the repulsive potential $V = Q/r$ dominates the local kinetic energy terms, ``classicalizing'' the energy landscape and shifting the spectral peak closer to the statistical mean. However, the fact that $R_E$ remains strictly bounded by $0.5$ underscores that 3D quantum spatial spread imposes an inescapable topological constraint on the fragment energy spectrum.

\subsubsection{The Shell-Dependent Detection Crisis}
The decay in $R_E$ is observed as the orbital exponent $\zeta$ increases (red curve in Fig.~\ref{fig:zeta_q_ratio}). Within our framework, $\zeta$ acts as a proxy for localization and effective nuclear charge; small $\zeta$ values characterize diffuse valence orbitals, while large $\zeta$ represent highly localized core-shell states. As $\zeta$ increases, the quadratic scaling of the quantum pressure near the coordinate origin drives $\langle E \rangle$ to values vastly exceeding $E_{\text{peak}}$, which remains constrained by the most probable radial position.

Our findings identify a \textbf{shell-dependent detection crisis}: while valence-shell fragmentation (low $\zeta$) remains relatively transparent to calorimetry, inner-shell events suffer from massive geometric masking. For $\zeta \approx 5$, a detector may resolve as little as 15\% of the true interaction energy. Given that localization typically scales with fragment mass, the extreme confinement of nucleons in nuclear physics implies a significantly suppressed $R_E$. The global contour map in Fig.~\ref{fig:coulomb_zeta_ratio} synthesizes these trends, confirming that the deeper a state is localized within the potential well, the more effectively 3D quantum geometry masks the true energy budget from classical detection.
\section{The Dimensionless Master Curve and Scaling Symmetry}

\begin{figure*}[htbp]
\centering
    \begin{subfigure}[b]{0.48\textwidth}
        \centering
        \includegraphics[width=\textwidth]{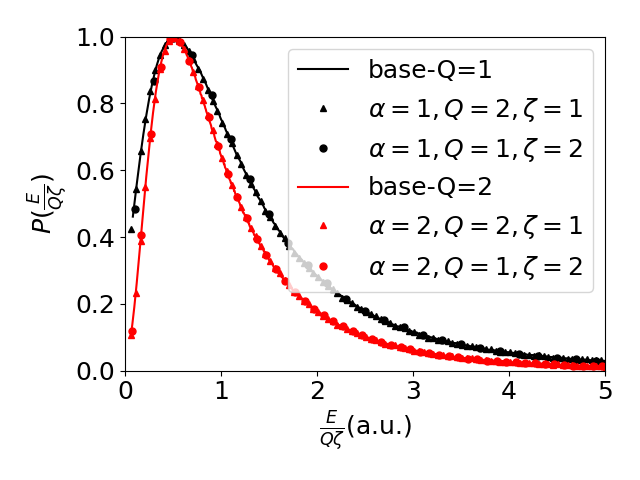}
        \caption{} 
        \label{fig:scaling_dis}
    \end{subfigure}
    \hfill
    \begin{subfigure}[b]{0.48\textwidth}
        \centering
        \includegraphics[width=\textwidth]{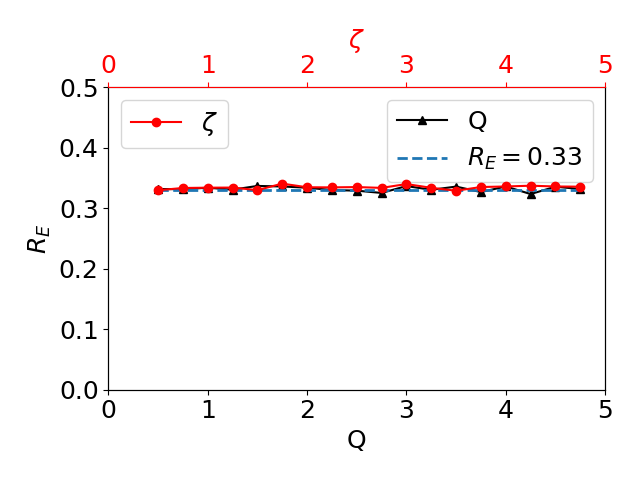}
        \caption{} 
        \label{fig:zeta_q_ratio_scaling}
    \end{subfigure}

    \vspace{10pt} 

    \begin{subfigure}[b]{0.48\textwidth}
        \centering
        \includegraphics[width=\textwidth]{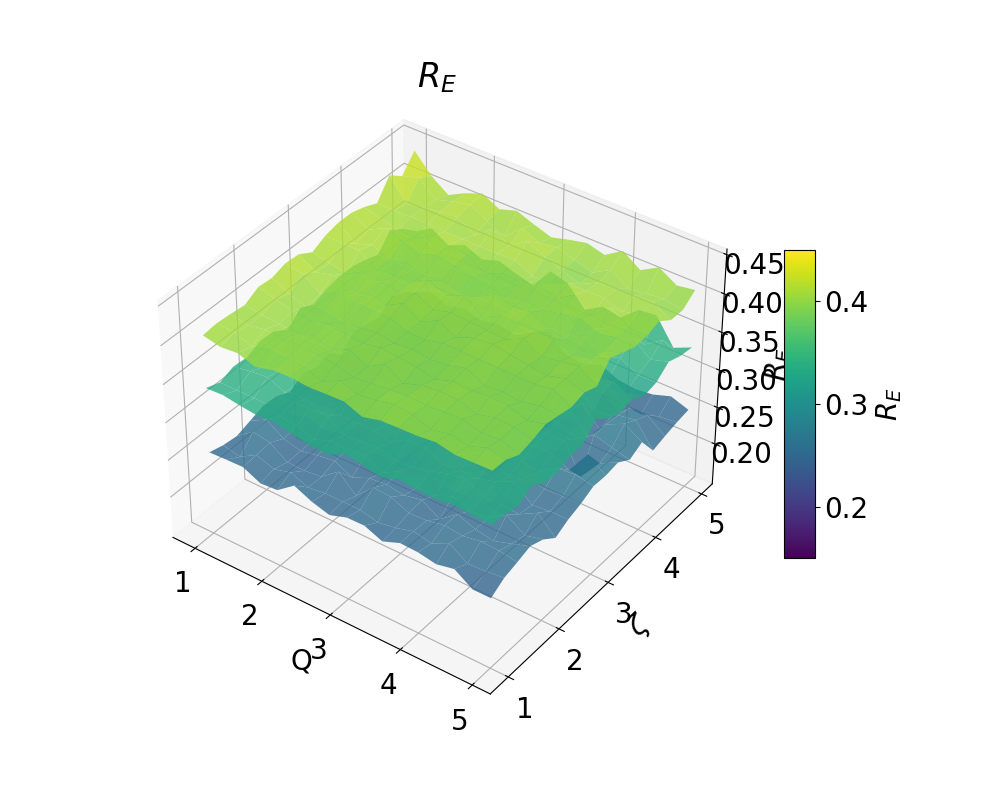}
        \caption{} 
        \label{fig:all_scaling}
    \end{subfigure}
    \hfill
    \begin{subfigure}[b]{0.48\textwidth}
        \centering
        \includegraphics[width=\textwidth]{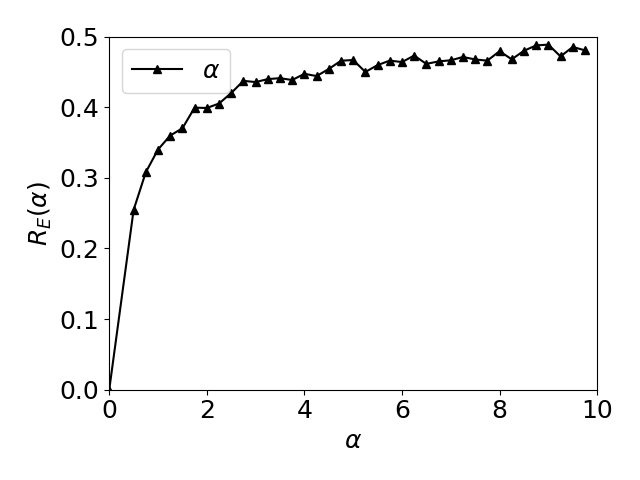}
        \caption{} 
        \label{fig:ratio_alpha}
    \end{subfigure}
    \caption{\textbf{Universal Scaling and the 0.33 Geometric Landmark.} 
        \textbf{Invariant scaling and the universal 0.33 geometric landmark.} 
        \textbf{(a)} Evolution of the peak-to-mean ratio $R_E$ as a function of repulsive charge $Q$ under the constraint $\zeta=Q, M=1, n=1$. The dashed horizontal line indicates the 0.33 stability limit. 
        \textbf{(b)} $R_E$ as a function of orbital exponent $\zeta$ for the case $Q=1, M=\zeta, n=1$, exhibiting identical convergence to the 0.33 landmark. 
        \textbf{(c)} Three-dimensional contour representation of the $(\zeta, Q)$ parameter space for $(\zeta, Q) \in [1, 5]$. The visualization shows three invariant planes corresponding to dimensionless scaling factors $\alpha = 0.5$ ($R_E \approx 0.26$), $\alpha = 1.0$ ($R_E \approx 0.33$), and $\alpha = 2.0$ ($R_E \approx 0.39$), calculated with the mass scaling $M = \alpha \zeta / Q$. 
        \textbf{(d)} Explicit mapping of the energy ratio $R_E$ as a function of the dimensionless coupling constant $\alpha$. The baseline computation uses parameters $M=\zeta=Q=n=1$.
    }
    \label{fig:master_scaling}
\end{figure*}

The emergence of the 0.33 landmark is rooted in a fundamental scaling symmetry inherent to three-dimensional fragmentation dynamics. To characterize this universality, we examine the peak-to-mean ratio $R_E$ across the multi-dimensional parameter space of mass ($M$), orbital localization ($\zeta$), and repulsive charge ($Q$). As demonstrated in Fig.~\ref{fig:master_scaling}, the energy morphology is governed by a single dimensionless coupling constant $\alpha$, defined as:
\begin{equation}\label{eq:scaling}
    \alpha = \frac{MQ}{\zeta}.
\end{equation}

For a $1s$ Slater-type orbital, substituting this scaling relation reveals that the local kinetic density scales as $\rho_{T} \propto \zeta^2/M = \alpha^{-1} \zeta Q$, while the local potential density follows $\rho_{V} \propto Q\zeta$ (see derivation in Sec.~\ref{sec:Derivation}). By transforming to a dimensionless radial coordinate $r_s = \zeta r$, we find that both energy components at every coordinate point scale linearly with the product $\zeta Q$ for any fixed $\alpha$. This \textit{local synchronization} ensures that the geometric morphology of the resulting $P(E)$ distribution is invariant to the absolute energy scale. This symmetry is confirmed in Fig.~\ref{fig:scaling_dis}, where normalized distributions $P(E/Q\zeta)$ for disparate $Q$ and $\zeta$ collapse onto identical master curves. With the $Q\zeta$ scaling of $E$, the distribution converges toward a fixed peak position, providing a robust, scale-invariant baseline for extracting intrinsic physical parameters from fragmented experimental spectra.

Furthermore, this symmetry implies that the detection ratio $R_E$ remains invariant along $M \propto \zeta$ trajectories. The system undergoes a self-similar expansion in energy space where the $4\pi r^2$ masking effect is preserved as a constant geometric partition of the total energy budget. The invariance observed in the parallel trajectories of Fig.~\ref{fig:zeta_q_ratio_scaling} and the invariant planes of Fig.~\ref{fig:all_scaling} is a direct consequence of this homogeneous scaling. At the $\alpha \approx 1$ equilibrium, this symmetry manifests as a 1D plateau at $R_E \approx 0.33$, identifying the landmark as a structural constant of 3D quantum geometry rather than a transient dynamical variable.

To determine the universal values of $R_E$ across the scaling parameter $\alpha$, we utilize this invariance to map the ratio onto our previously computed parametric results:
\begin{equation}\label{eq:RE-equiva}
    R_E(\alpha) = R_E(Q)\rvert_{M=\zeta=1}.
\end{equation}
As shown in the master curve in Fig.~\ref{fig:ratio_alpha}, the numerical behavior of $R_E$ is characterized by the following asymptotic limits:
\begin{equation}\label{eq:RE-limit}
    \lim_{\alpha\to 0} R_E = 0, \quad \lim_{\alpha\to \infty} R_E = 0.5.
\end{equation}

At $\alpha=1$, the system achieves a \textbf{geometric equilibrium} where the $4\pi r^2$ volume element sequesters exactly two-thirds of the total energy in the high-momentum radial tail, yielding the 0.33 landmark. As $\alpha$ increases (representing high repulsion $Q$ or low mass $M$), the ratio monotonically approaches the $0.5$ universal bound—a regime where the repulsive field dominates the initial-state quantum pressure. 

Conversely, as the system moves toward high localization $\zeta$ (decreasing $\alpha$), $R_E$ decays precipitously, falling below $0.2$ for $\zeta > 5$. This signals a profound \textbf{detection crisis}: the vast majority of interaction energy becomes experimentally invisible to peak-centric calorimetry. This confirms that for sufficiently localized states—such as core-shell electrons or compressed nucleons—the true interaction energy is effectively masked by the topological constraints of 3D space. Consequently, the mapping $R_E(\alpha)$ serves as a universal master curve for energy reconstruction across all scales of three-dimensional quantum fragmentation.

\section{Discussion: From Molecular Calorimetry to Subatomic Decay}

\begin{figure}[htbp]
    \centering
\begin{subfigure}[b]{0.44\textwidth}
        \centering
        \includegraphics[width=\textwidth]{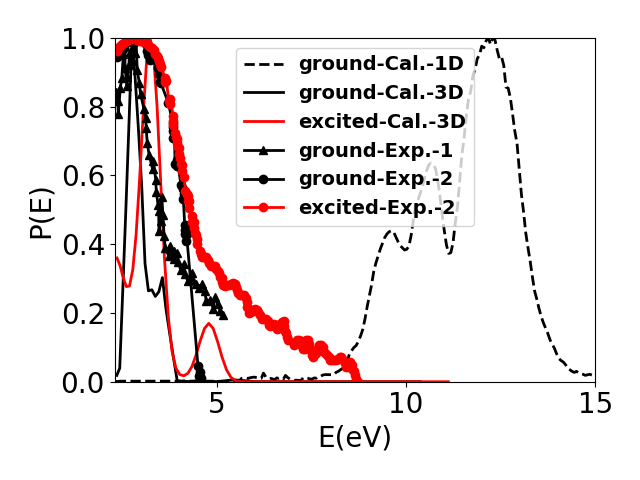}
        \caption{}
        \label{fig:h2p_comp}
    \end{subfigure}
    \hfill
    \begin{subfigure}[b]{0.40\textwidth}
        \centering
        \includegraphics[width=\textwidth, clip]{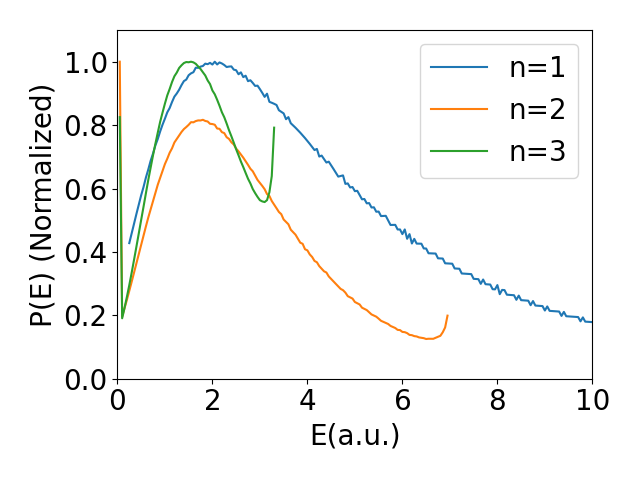}
        \caption{}
        \label{fig:n_dis}
    \end{subfigure}
    \hfill
    \begin{subfigure}[b]{0.40\textwidth}
        \centering
        \includegraphics[width=\textwidth]{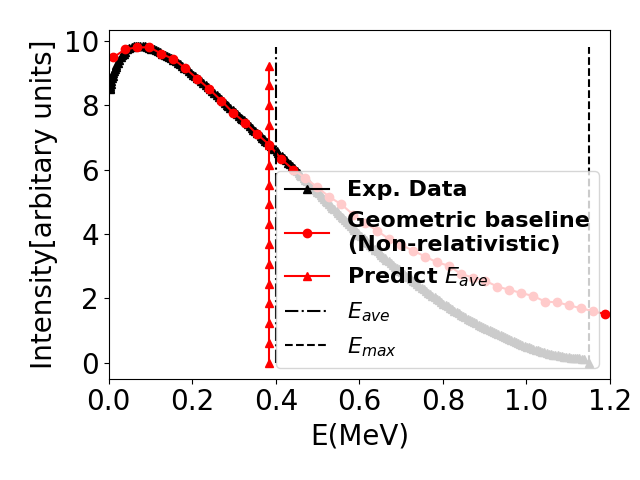}
        \caption{}
        \label{fig:beta_decay}
    \end{subfigure}
\caption{
\textbf{Energy distribution analysis and empirical benchmarks.} 
\textbf{(a)} Proton KER spectra for $H_2^+$ dissociative ionization. Numerical results represent 1D (dashed lines, Ref.~\cite{Yue2013}) and 3D (solid lines, this work) computations. Experimental data are denoted by triangle markers (Ref.~\cite{Pavicic2005}) and dot markers (Ref.~\cite{Rudenko_2005}, after applying a logarithmic transformation). Black curves correspond to ground-state ($n=1$) contributions; red curves include the summation of populated excited states.
\textbf{(b)} Energy distribution $P(E)$ as a function of principal quantum number $n$ ($n=1, 2, 3$) for fixed $Q=1, \zeta=1$. 
\textbf{(c)} Comparison of the historical $^{210}$Bi beta-decay spectrum (triangle markers, Ref.~\cite{Ellis1927}) with the theoretical $P(E)$ distribution at $\alpha=0.16$ (red dotted line). 
}
\label{fig:exprimental-data}
\end{figure}

\subsection{Empirical Validation: Resolving the $H_2^+$ Energy Anomaly}
The physical validity of the 0.33 geometric landmark is most strikingly evidenced by the long-standing energy discrepancy in attosecond $H_2^+$ fragmentation. As shown in Fig.~\ref{fig:h2p_comp}, traditional 1D semiclassical models~\cite{Yue2013} predict a kinetic energy release (KER) peak between $10$--$13$~eV (black dashed line), significantly overestimating experimental observations (black triangles). In contrast, our 3D formulation yields a primary peak near $3$~eV (black solid line), in excellent agreement with experimental calorimetry (black triangles)~\cite{Pavicic2005,Rudenko_2005}.

Our model replicates the observed $\approx 7$~eV shift without invoking external field corrections or complex multi-photon dynamics, identifying the $4\pi r^2$ volume element as the primary driver of the discrepancy. While the large mass of the escaping protons ($M_p \approx 1836$~a.u.) would classically drive the system toward a high $R_E$ ratio, their initial localization within the $1\sigma_g$ molecular manifold provides a critical \textit{symmetry-induced compensation}. Because the protons are constrained along the internuclear axis—a distribution templated by the axial $\psi \propto r \cos^m \theta e^{-\zeta r}$ symmetry—the angular components of the Laplacian ($\nabla^2 \psi \propto m/r^2$) significantly enhance the local ``quantum pressure.'' This effect effectively neutralizes the mass scaling in the dimensionless coupling $\alpha = MQ/\zeta$, maintaining the heavy fragment system near the $\alpha \approx 1$ geometric equilibrium. Consequently, the detected proton energy represents only one-third of the integrated mean ($\langle E \rangle \approx 11$~eV), confirming that the ``missing energy'' in molecular calorimetry is a direct manifestation of 3D quantum topology.

\subsection{Nodal Structures and Spectral Broadening}
Beyond the ground state ($n=1$), higher principal quantum numbers introduce radial nodes that alter the energy distribution $P(E)$. As illustrated in Fig.~\ref{fig:n_dis}, simulations for $n=1, 2,$ and $3$ reveal that while the primary low-energy peak remains stable, increasing $n$ triggers secondary high-energy features. Physically, the local kinetic energy $T_{\text{local}}(r)$ incorporates terms related to the curvature of these additional nodes, redistributing probability density into higher kinetic energy regimes and creating a multi-modal spectrum.

This nodal redistribution provides a geometric explanation for the spectral broadening observed in high-intensity $H_2^+$ experiments (red and black dots in Fig.~\ref{fig:h2p_comp}). When excited states ($n > 1$) are populated, our 3D model identifies secondary features between $4$--$5$~eV (red solid curve in Fig.~\ref{fig:h2p_comp}) which are absent in 1D approximations. These features serve as a direct signature of the radial nodal structures in the initial-state wavepacket. Thus, while the $1/3$ landmark defines the primary detection peak, the population of higher-order shells provides the underlying geometric mechanism for spectral width evolution.

\subsection{The Cascading Geometric Mask: A Two-Process Model of $\beta$ Decay}

The identification of the $\alpha = MQ/\zeta$ scaling law provides a provocative lens for interpreting energy anomalies in nuclear physics. Historically, the continuous electron spectrum in $\beta$ decay---characterized by a peak and average energy significantly below the endpoint ($\langle E \rangle \approx 1/3 E_{\text{max}}$)---was the catalyst for the neutrino hypothesis~\cite{Ellis1927}. While our model replicates the spectral morphology using a coupling of $\alpha=0.16$ in the non-relativistic regime(Fig.~\ref{fig:beta_decay}), we propose that the total energy deficit arises from a \textbf{cascading geometric mask} involving two distinct quantum processes.

In this framework, we consider the fragmentation as a sequential interaction between a primary nuclear fragment (particle $A$) and a secondary inner-shell electron (particle $B$). We characterize the event through the following hierarchy:

\begin{enumerate}
    \item The Geometric Gate (Process 1): Particle $A$ is born from a highly localized nuclear environment. Before interacting with the atomic shell, its own energy distribution is subject to the 3D geometric baseline. We assume that the effective interaction energy $A$ provides to the system is constrained by its own spectral peak ($E_{\text{peak}} \approx 1/3 Q$, marked by red trigangles in Fig.~\ref{fig:beta_decay}). This pre-filtering defines the maximum energy budget available for secondary ejections.
    
    \item Potential Inversion and Overlap (Process 2): As particle $A$ echelons toward the atomic shell, particle $B$ experiences a sudden transition from an attractive to a repulsive potential ($V \to +Q$). The transfer of energy from $A$ to $B$ is governed by the quantum overlap integral $\langle \psi_B | \hat{V} | \psi_A \rangle$. 
\end{enumerate}

Crucially, the interaction is dominated by regions of maximum spatial density. Because both $A$ and $B$ are 3D wavefunctions filtered by the $4\pi r^2$ volume element, their maximum overlap occurs precisely at their respective radial peaks. This implies a **Quantum Selection Rule**: only the peak-energy component of particle $A$'s wavepacket can effectively couple with particle $B$.

The remaining two-thirds of the total nuclear energy, sequestered in the high-curvature radial tails of the wavefunctions, remains topologically inaccessible to the secondary electron. Consequently, the detected $\beta$ spectrum is a convolution of two geometric constraints. This explains why the observed energy is tethered to the 0.33 landmark not as a result of weak-force kinematics, but as a structural necessity of 3D wave-coupling. By mapping these decay modes onto our universal master curve, we demonstrate that the "missing energy" in localized subatomic systems is a direct manifestation of this cascading geometric sequestration.

\subsection{Scaling Limits: Comparing Alpha and Beta Decay}
The universal nature of the scaling law is further underscored by the contrasting morphologies of $\alpha$ and $\beta$ decay. While the low-mass regime of $\beta$ decay resides near the $\alpha \approx 1$ landmark—where geometric masking is maximal—the emission of heavy helium nuclei in $\alpha$ decay represents the high-$\alpha$ limit. As $M$ increases, our model predicts a geometric sharpening of $P(E)$ and a convergence of $R_E$ toward the classical limit. This explains why $\alpha$ spectra appear as nearly mono-energetic lines with high detection efficiency, in contrast to the broad distributions of $\beta$ decay. The shift from masked to visible energy peaks confirms that the observability of subatomic transitions is fundamentally governed by the dimensionless scaling of the 3D volume element.

\subsection{Orbital Contraction and Relativistic Scaling}
The spectral distribution $P(E)$ is critically governed by the localization parameter $\zeta$. In high-$Z$ nuclei, the proximity of inner-shell electrons effectively scales $\zeta$, leading to the spectral broadening observed in traditional nuclear studies~\cite{Wilkinson1990}. For superheavy nuclei ($Z > 100$), extreme localization increases the probability of detecting ultra-high energy fragments, a phenomenon consistent with experimental observations of relativistic orbital contraction~\cite{Henning2012}. While currently interpreted via Fermi theory, our model suggests these shifts are consistent with the geometric pressure inherent to high-$\zeta$ localized states, providing a scale-invariant link between atomic and relativistic nuclear dynamics.
\section{Conclusion}

In this work, we have demonstrated that the persistent discrepancy between theoretical energy budgets and experimental peaks in quantum fragmentation is not a consequence of technical limitations, but a fundamental topological artifact of three-dimensional space. By synthesizing localized Slater-type orbitals with full-dimensional TDSE benchmarks, we have established that the $4\pi r^2$ volume element acts as an inescapable geometric filter. This mechanism inherently masks a significant portion of a system's interaction energy, imposing a universal peak-to-mean bound of $R_E < 0.5$ on all 3D fragmentation spectra.

The discovery of the invariant scaling law, $\alpha = MQ/\zeta$, provides a unified framework for energy reconstruction across twelve orders of magnitude. We have shown that the 7~eV ``missing energy'' in attosecond $H_2^+$ fragmentation is a direct manifestation of this geometric constraint, resolving a long-standing anomaly without the need for external field corrections or complex dynamical modeling. Most significantly, the alignment between the 0.33 landmark and the historical energy morphology of $\beta$ decay suggests that the $1/3$ ratio is a universal baseline for energy detection in localized 3D systems.

Our findings provide a transformative perspective on the interpretation of energy anomalies in subatomic physics. By mapping disparate decay modes onto a universal master curve, we demonstrate that the broad spectrum of $\beta$ decay and the high-resolution efficiency of $\alpha$ decay represent two limits of the same geometric scaling. While the neutrino remains a fundamental component of the Standard Model, our results reveal that the baseline for energy detection in localized states is inherently suppressed by the topology of the initial-state wavefunction. 

Because localization ($\zeta$) correlates with particle mass ($M$), subatomic systems are naturally driven toward regimes where the detection ratio $R_E$ is highly sensitive to 3D masking. Consequently, these geometric effects must be rigorously decoupled from signatures of invisible flux or new particles. This work shifts the focus of future calorimetry from the search for undetected mass to the precision calibration of the intrinsic geometric filters inherent to our three-dimensional quantum world.

\appendix

\section{Methods}

\subsection{Sudden Approximation and Hamiltonian Transmutation}\label{sec:sudden}

The applicability of our 3D fragmentation model to high-energy decay processes is justified via the \textit{sudden approximation}. In attosecond molecular physics, the rapid ionization of bonding electrons transforms an attractive potential into a repulsive landscape before the nuclear wavefunctions can adiabatically adjust. 

This framework can be extended to the nuclear regime, where reaction energies are sufficiently high to ensure that emitted particles reach ultra-relativistic velocities before the surrounding electronic environment can respond. In $\beta$-decay, for instance, the transmutation timescale is significantly shorter than the characteristic orbital period of the inner-shell electrons. Consequently, the electronic wavefunctions remain spatially invariant (or "frozen") during the event. The newly born high-speed electron introduces a localized Coulombic repulsion that suddenly competes with the nuclear attraction. This process, analogous to the Migdal "shake-off" or Auger-like transitions, forces a reconfiguration of the electronic cloud.

\subsection{Bohmian Local Energy Derivation}
The local kinetic energy is derived from the Laplacian of the initial spatial wavefunction: $T_{local}(\mathbf{r}) = -\frac{1}{2M} \frac{\nabla^2 \psi_0(\mathbf{r})}{\psi_0(\mathbf{r})}$. In spherical coordinates, this yields the radial dependence utilized in our $P(E)$ construction. Because each Bohmian particle conserves its total energy in a time-independent repulsive field, the distribution at $t=0$ is identical to the asymptotic kinetic energy distribution at $t \to \infty$.

\subsection{Numerical Validation against TDSE}\label{sec:fitting}
\begin{figure*}[htbp]
    \centering
    \begin{subfigure}[b]{0.32\textwidth}
        \includegraphics[width=\textwidth, clip]{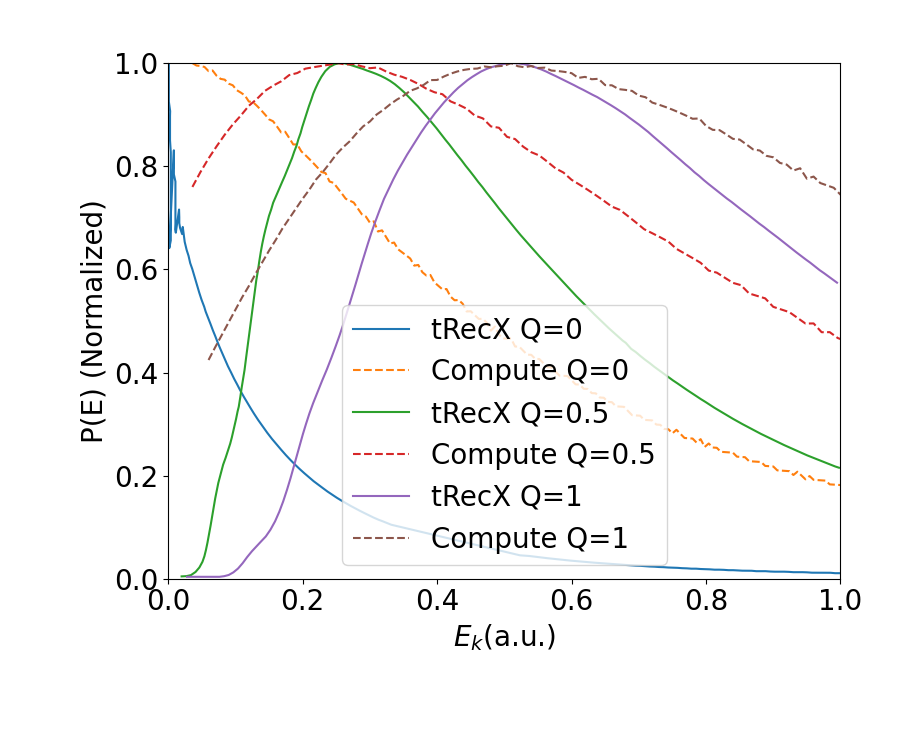}
        \caption{} 
        \label{fig:trecx_compare}
    \end{subfigure}
    \begin{subfigure}[b]{0.32\textwidth}
        \includegraphics[width=\textwidth, clip]{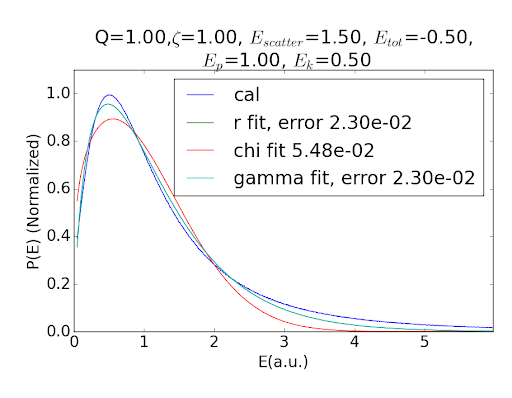}
        \caption{} 
        \label{fig:fit-1}
    \end{subfigure}
    \begin{subfigure}[b]{0.32\textwidth}
        \includegraphics[width=\textwidth, clip]{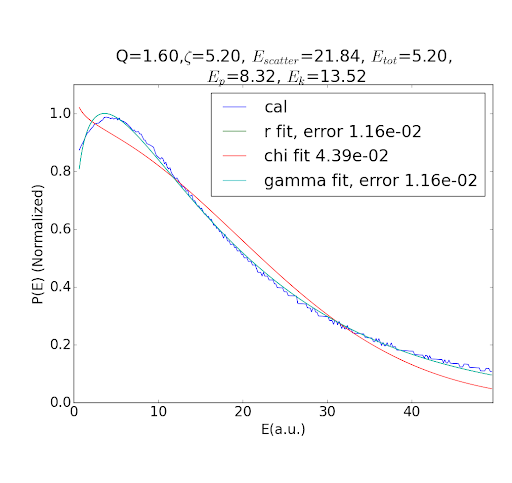}
        \caption{} 
        \label{fig:fit-2}
    \end{subfigure}
    \caption{\textbf{Energy Distribution Analysis.} 
        \textbf{(a)} Comparison of normalized energy distributions $P(E)$ between the present model (solid) and tRecX TDSE simulations (dashed) for $Q=0, 0.5$ ($\zeta=1$). 
        \textbf{(b)} Analytical fitting for $\zeta=1, Q=1$. 
        \textbf{(c)} Results for $\zeta=5.2, Q=1.6$. The r-fit (green) and Gamma fit (cyan) identify the radial nature of the distribution. Note that $E_{peak}$ remains significantly lower than $E_{scatter}$, consistent with the expected energy reduction for higher $\zeta$ values in nuclear repulsion scenarios.}
        \label{fig:interpolate}
\end{figure*}

To validate the predictive power of the 3D model, we compared $P(E)$ against full quantum simulations performed via the \texttt{tRecX} TDSE package~\cite{SCRINZI2022108146}.
Using a 1s hydrogenic orbital ($\zeta=1$ and $l=0$), we found that peak positions show high consistency across all repulsive charges  $Q=0,0.5,1$ (see Fig.~\ref{fig:trecx_compare}).
Deviations in the distribution tails are attributed to the numerical constraints of grid-based solvers at $r \to 0$, whereas our analytical approach maintains precision in the near-nucleus region.

To identify a universal analytical form for the fragment energy spectrum, we performed a regression analysis on the energy distribution $P(E)$ using several statistical models. Fig.~\ref{fig:interpolate} illustrates the results for a ground-state hydrogenic orbital ($\zeta=1,Q=1$, Fig.~\ref{fig:fit-1}) and a more localized, higher-charge configuration ($\zeta=5.2,Q=1.6$, Fig.~\ref{fig:fit-2}). We compared the computed data (blue) against three candidate distributions: a radial-type distribution (r-fit), a Gamma distribution, and a $\chi^2$ distribution. As shown in in the figure, the r-fit (green) and the Gamma distribution (cyan) produce the highest fidelity, with a residual error of less than 2\%. The near-perfect overlap between these two functions confirms that the energy distribution essentially obeys a radial distribution law.
\subsection{Symmetry-Dependent Local Kinetic Energy}

To evaluate the influence of orbital alignment on the energy ratio, we consider a generalized prolate orbital often encountered in molecular ionization:
\begin{equation}
    \psi(r, \theta) = \mathcal{N} r \cos^m \theta \exp(-\zeta r)
\end{equation}
where $n$ represents the degree of axial alignment. The Laplacian in spherical coordinates, $\nabla^2 = \frac{1}{r^2}\frac{\partial}{\partial r}(r^2\frac{\partial}{\partial r}) + \frac{1}{r^2 \sin \theta}\frac{\partial}{\partial \theta}(\sin \theta \frac{\partial}{\partial \theta})$, applied to this state yields:
\begin{equation}
    \frac{\nabla^2 \psi}{\psi} \approx \left( \zeta^2 - \frac{4\zeta}{r} + \frac{2(1-m)}{r^2} \right)
\end{equation}
The local kinetic energy $T_{\text{local}}(r) = -\frac{1}{2M}\frac{\nabla^2 \psi}{\psi}$ therefore incorporates a centrifugal-like term proportional to the alignment $m$. In $H_2^+$, the high value of $n$ effectively counteracts the large mass $M$ in the denominator, preventing the collapse of the energy distribution into the classical regime. This confirms that the 0.33 landmark is a robust attractor for both light spherical systems and heavy aligned systems.

\subsection{Derivation of the Universal Scaling Law}\label{sec:Derivation}

To establish the physical basis for the 0.33 landmark, we examine the local energy densities of a system described by a $1s$ Slater-type orbital:
\begin{equation}
\psi(r) = \sqrt{\frac{\zeta^3}{\pi}} e^{-\zeta r}
\end{equation}
The detected kinetic energy release (KER) is determined by the local mapping of potential $V(r)$ and kinetic $T(r)$ energy densities onto the $P(E)$ distribution, weighted by the 3D volume element $dV = 4\pi r^2 dr$. By transforming to a dimensionless scaled radial grid $r_s = \zeta r$, we reveal the universal scaling properties of the fragmentation morphology.

\subsubsection*{1. Potential Energy Scaling}
For a repulsive Coulombic interaction $V(r) = Q/r$, the local potential energy density $\rho_V(r) = \psi^2(r) V(r) 4\pi r^2 dr$ is expressed in the scaled grid as:
\begin{equation}
\rho_V(r_s) = 4\zeta Q \cdot r_s e^{-2r_s} dr_s
\end{equation}
This confirms that at every point on the universal grid $r_s$, the potential energy density scales as $\rho_V(r_s) \propto Q\zeta$.

\subsubsection*{2. Kinetic Energy Scaling}
The radial kinetic energy density is derived from the Laplacian operator $\hat{T} = -\frac{1}{2M} \nabla^2$. Acting on the $1s$ orbital, the operator yields:
\begin{equation}
\hat{T}\psi(r) = -\frac{1}{2M} \left( \frac{d^2}{dr^2} + \frac{2}{r}\frac{d}{dr} \right) \psi(r) = -\frac{\zeta^2}{2M} \left( 1 - \frac{2}{\zeta r} \right) \psi(r)
\end{equation}
Transforming the kinetic density $\rho_T(r) = \psi(r) \hat{T} \psi(r) 4\pi r^2$ to the scaled grid $r_s$ yields:
\begin{equation}
\rho_T(r_s) = -\frac{2\zeta^2}{M} \left( 1 - \frac{2}{r_s} \right) r_s^2 e^{-2r_s} dr_s
\end{equation}
In this coordinate system, the local kinetic density scales as $\rho_T(r_s) \propto \zeta^2/M$. By introducing the dimensionless coupling constant $\alpha = MQ/\zeta$, we substitute $M = \alpha \zeta / Q$ to obtain:
\begin{equation}
\frac{\zeta^2}{M} = \frac{\zeta^2}{\alpha \zeta / Q} = \frac{1}{\alpha} Q\zeta
\end{equation}

\subsubsection*{3. Homogeneous Invariance}
At the geometric equilibrium defined by $\alpha = 1$, both potential and kinetic energy densities at every scaled grid point $r_s$ scale identically with the product $Q\zeta$:
\begin{equation}
\rho_V(r_s) \propto (Q\zeta) \cdot f(r_s), \quad \rho_T(r_s) \propto (Q\zeta) \cdot g(r_s)
\end{equation}
Since the ratio $\rho_V(r_s) / \rho_T(r_s)$ is independent of the absolute values of $Q$ and $\zeta$, the morphology of the resulting $P(E)$ distribution is self-similar. This mathematical homogeneity ensures that the detection ratio $R_E = E_{\text{peak}}/\langle E \rangle$ remains locked at the 0.33 landmark, manifesting as the invariant horizontal plateaus observed in our parameter space analysis.

\subsection{$H_2^+$ Simulation Protocols and Parameter Selection}

To validate our geometric model, full quantum-dynamical simulations of $H_2^+$ were performed using the \texttt{tRecX} software package. These simulations solve the three-dimensional TDSE to extract high-fidelity KER spectra.

Two distinct ionization regimes were utilized to benchmark the influence of initial-state geometry: \begin{enumerate} \item \textbf{Ground-State Manifold:} Simulations using linearly polarized laser fields were employed to isolate the ground-state dissociation pathway. \item \textbf{Excited-State Populations:} Circularly polarized fields were utilized to drive enhanced population transfer into higher-lying electronic states (n>1). This protocol aligns with both experimental and theoretical findings~\cite{Zhangetal2023, Geng2022StrongField, Rudenko_2005}, which demonstrate that circular polarization significantly broadens the KER spectrum by populating excited-state manifolds. \end{enumerate}

Consistent with established experimental conditions~\cite{Pavicic2005}, the laser parameters were set to a central wavelength of 800~nm with a peak intensity of $8\times 10^{13}W/cm^2$. The resulting 3D flux was integrated to produce the KER distributions compared in the lower panel of Fig.~\ref{fig:exprimental-data}.

\bibliography{h2plus.bib}
\end{document}

%% file: my_commands.tex
\usepackage{amsmath}
\usepackage{amssymb}
\usepackage{makeidx}
\usepackage{graphicx}
\usepackage{color}

\newcommand{\hide}[1]{}

\newcommand{\lcase}{\left\{\begin{array}{ll}}
\newcommand{\rcase}{\end{array}\right.}
\newcommand{\ear}{\end{array}}
\newcommand{\bal}{\begin{align}}
\newcommand{\eal}{\end{align}}
\newcommand{\bma}{\begin{pmatrix}}
\newcommand{\ema}{\end{pmatrix}}
\newcommand{\beq}{\begin{equation}}
\newcommand{\eeq}{\end{equation}}
\newcommand{\bel}[1]{\begin{equation}\label{eq:#1}}
\newcommand{\eel}{\end{equation}}
\newcommand{\bea}{\begin{eqnarray}}
\newcommand{\eea}{\end{eqnarray}}
\newcommand{\beaNN}{\begin{eqnarray*}}
\newcommand{\eeaNN}{\end{eqnarray*}}

\newcounter{lecture}

